\documentclass[epj,final]{svjour}
\usepackage{amsmath,bm,graphicx,color}

\newcommand{\bra}[1]{\langle{#1}|}
\newcommand{\ket}[1]{|{#1}\rangle}

\begin{document}

\title{Robust stationary entanglement of two coupled qubits in
independent environments}

\author{M. Scala\inst{1}
\and R. Migliore\inst{2}
\and  A. Messina\inst{1}
\and L.L. S\'anchez-Soto\inst{3}}

\institute{Dipartimento di Scienze Fisiche ed
  Astronomiche dell'Universit\`{a} di Palermo,
 via Archirafi 36, 90123  Palermo, Italy
\and
CNR, CNISM, via Archirafi 36, 90123
 Palermo, Italy
\and
Departamento de \'Optica, Facultad de F\'isica,
Universidad Complutense, 28040 Madrid, Spain}

\date{Received: \today, Revised version: date}

\abstract{
  The dissipative dynamics of two interacting qubits coupled to
  independent reservoirs at nonzero temperatures is
  investigated, paying special attention to the entanglement
  evolution. The counter-rotating terms in the
  qubit-qubit interaction give rise to stationary entanglement,
  traceable back to the ground state structure.  The robustness
  of this entanglement against thermal noise is thoroughly
  analyzed, establishing that it can be detected at reasonable
  experimental temperatures.  Some effects linked to a possible
   reservoir asymmetry are brought to light.
}

\maketitle

\section{Introduction}

The entanglement dynamics of open quantum systems has attracted
considerable attention over recent years. Quantum correlations are
indeed central in a variety of contexts, both because of their
fundamental properties and because they are key ingredients in the
fields of quantum communication, computation and information
processing.

Superconducting devices are among the best candidates for {quantum
state engineering}.  The nonlinearity of Josephson junctions can be
used to mimic two-level quantum systems (qubits), with the
advantages that the inherently low dissipation of superconductors
make long coherence times possible, while integrated circuit
technology allows for scaling to large and complex
systems~\cite{makhlin,noritoday,devoret}.

Recent experiments {have} successfully demonstrated the existence of
quantum coherent oscillations in individual~\cite{1,2,3,4,5} and
coupled~\cite{6,7,8,9,10} Josephson qubits. Yet many aspects of
decoherence and dissipation in these systems still remain to be
understood. A current challenge is to discriminate among different
sources of decoherence. This calls for finding appropriate
representations of the interaction between these devices and the
external environment, so as to single out the conditions (quite
often model-dependent) under which a coherent behavior is still
visible.

Within this scenario, the simple model of two coupled {qubits} in
thermal contact with different baths is more than an academic
curiosity, {for} it can describe quite different experimental setups
as, for instance, two inductively (capacitively) coupled flux
(charge) qubits, provided that their spatial separation is large.
{It also} allows one to investigate the mechanisms that determine
the degradation of coherent Rabi oscillations in Josephson phase
qubits stemming from spurious microwave resonators. These
resonators, as shown by Martinis and coworkers~\cite{martinis},
arise from changes in the junction critical current produced by
two-level states in the tunnel barrier. {One} bath describes {then}
the electromagnetic environment of the phase qubit, while the other
one mimics the phononic environment coupled to the junction
microwave resonator. Finally, such an {approach} is of relevance in
other frameworks, such as cavity
QED~\cite{yonac,yonac2,lopez,yonac3}, quantum dots, and spin
systems~\cite{dot,amico}.

Quite recently, we solved the zero-temperature dynamics of a model
that includes the usually omitted counter-rotating terms in the
qubit-qubit interaction. Besides confirming many {features
previously reported} in the
literature~\cite{Yu,bellomo,zoller,storcz2,Tanas,tanas 2008,grifoni,storcz,quiroga,sinaysky,itakura,sina2,huelga1,huelga2}, our
analysis showed that these terms {produce} long-time stationary
entanglement~\cite{JPHYA,Ustron,Nori2009}. The next natural task is thus to
look for effects stemming from nonzero reservoir temperatures.

In this {vein}, we {work out here} the solution of the master
equation at generic reservoir temperatures, {establishing} that the
effects of the counter-rotating terms turn out to be very robust
against thermal noise, since {they prevail} at temperatures at which
Josephson qubits work in actual experiments.  We put forward the
role of the qubit-qubit coupling constant $\lambda$ in the creation
of stationary entanglement at finite temperatures: the optimum case
can be reached for values of $\lambda$ of the order of the qubit
Bohr frequencies, a condition easily {met in
  practice}. We finally show how these results are modified by
possible asymmetries between the reservoirs.

The paper is structured as follows. The decay model, the dynamics at
zero temperature and the comparison with the results for a single
common reservoir are presented and discussed in Sec.II. The solution
at generic reservoir temperatures and the analysis of the entanglement
dynamics is given in Sec.~III for different initial conditions. We
also investigate the effects of different reservoir temperatures and
qubit-reservoir coupling strengths. Finally, our conclusions are
summarized in Sec.~IV.

\section{The decay model and the dynamics at zero temperature}

\subsection{The model}

The system under study consists of a pair of coupled qubits.  Denoting
by $\ket{0}_1$ ($\ket{0}_2$) the ground state of the first (second)
qubit and by $\ket{1}_1$ ($\ket{1}_2$) the corresponding excited
state, the Hamiltonian is given by (with $\hbar =1$
throughout)~\cite{JPHYA}:
\begin{eqnarray}
  \label{Hamiltonianmodel}
  H_S = \omega_1 \, \sigma_+^{(1)}\sigma_-^{(1)} +
  \omega_2 \, \sigma_+^{(2)}\sigma_-^{(2)}
  + \frac{\lambda}{2} \,\sigma_x^{(1)}\sigma_x^{(2)},
\end{eqnarray}
$\omega_i$ being the Bohr frequency of the $i$th qubit and $\lambda$
the coupling constant, which, in the case of flux qubits with
flux-flux coupling, is proportional to their mutual
inductance~\cite{JJRosanna}. The Pauli operators are given by
$\sigma_+^{(i)}= | 1 \rangle_i \, {}_i \langle 0 |$, $\sigma_-^{(i)}=
| 0 \rangle_i \, {}_i \langle 1|$ and $\sigma_x^{(i)}= \sigma_+^{(i)}
+ \sigma_-^{(i)}$, with $i=1,2$.  Note that (\ref{Hamiltonianmodel})
contains counter-rotating interaction terms.

In reference~\cite{JPHYA}, a master equation was derived for the
case in which each qubit interacts with an independent bosonic
thermal bath.  Indeed, by using the general formalism given in
reference~\cite{petruccionebook}, under the secular approximation
and the assumption of {\em independent bosonic reservoirs}, the
Markovian master equation governing the dynamics of the qubit-qubit
system turns out to be:
\begin{eqnarray}
  \label{Meqfinale}
  & & \dot{\varrho}(t) = - i [ H_S, \varrho(t) ] \nonumber \\
  & & + c_I ( \ket{a} \bra{b} \, \varrho(t) \, \ket{b}\bra{a}-
   \frac{1}{2} \{ \ket{b} \bra{b}, \varrho(t) \} )\nonumber\\
  &&+ c_{II} (\ket{a} \bra{c} \varrho(t) \ket{c} \bra{a}-
  \frac{1}{2} \{ \ket{c} \bra{c}, \varrho(t) \} ) \nonumber\\
  & & + c_I (\ket{b} \bra{d} \, \varrho(t) \, \ket{d} \bra{b} -
  \frac{1}{2} \{ \ket{d} \bra{d }, \varrho(t) \} )\nonumber\\
  && +c_{II} ( \ket{c} \bra{d} \, \varrho(t) \, \ket{d} \bra{c} -
  \frac{1}{2} \{ \ket{d} \bra{d}, \varrho(t) \} ) \nonumber \\
  & & + \bar{c}_I ( \ket{b} \bra{a} \, \varrho(t) \, \ket{a}\bra{b} -
  \frac{1}{2} \{\ket{a}\bra{a}, \varrho(t) \} )\nonumber\\
  && +\bar{c}_{II} (\ket{c}\bra{a} \, \varrho(t) \, \ket{a}\bra{c} -
  \frac{1}{2} \{ \ket{a}\bra{a}, \varrho(t) \} )\nonumber\\
  & & + \bar{c}_I (\ket{d} \bra{b} \, \varrho(t) \,\ket{b} \bra{d} -
  \frac{1}{2} \{ \ket{b} \bra{b}, \varrho(t) \} )\nonumber\\
  &&+\bar{c}_{II} ( \ket{d} \bra{c} \, \varrho(t) \, \ket{c} \bra{d} -
  \frac{1}{2} \{ \ket{c} \bra{c}, \varrho(t) \} ) \nonumber\\
  & & + c_{\mathrm{cr},I} ( \ket{a} \bra{b} \, \varrho(t) \, \ket{d}\bra{c} +
  \ket{c} \bra{d} \, \varrho(t) \, \ket{b}\bra{a} ) \nonumber\\
  & & +
  c_{\mathrm{cr},II} ( \ket{a} \bra{c} \, \varrho(t) \, \ket{d}\bra{b} +
  \ket{b} \bra{d} \, \varrho(t) \, \ket{c} \bra{a}  ) \nonumber \\
  & & + \bar{c}_{\mathrm{cr},I}  ( \ket{d} \bra{c} \, \varrho(t) \, \ket{a}\bra{b} +
  \ket{b} \bra{a} \, \varrho(t) \,  \ket{c} \bra{d} )\nonumber\\
  && +
  \bar{c}_{\mathrm{cr},II} ( \ket{d} \bra{b} \, \varrho(t) \, \ket{a} \bra{c} +
  \ket{c}\bra{a} \, \varrho(t) \, \ket{b}\bra{d} ) \, .
\end{eqnarray}
The states appearing in the master equation, namely the eigenstates
of the Hamiltonian (\ref{Hamiltonianmodel}), {are given}, in the
uncoupled basis
$\left\{\ket{00},\ket{11},\ket{10},\ket{01}\right\}$, by
\begin{eqnarray}
  \label{eigenstates}
  \ket{a} & = &
  \cos  \frac{\theta_I}{2}  \ket{00} -
  \sin  \frac{\theta_I}{2}  \ket{11} \, ,
  \nonumber \\
  \ket{b}  & = & \cos  \frac{\theta_{II}}{2}  \ket{10} -
  \sin  \frac{\theta_{II}}{2}  \ket{01} \, ,
 \nonumber \\
  \ket{c} & = & \sin  \frac{\theta_{II}}{2}  \ket{10} +
  \cos  \frac{\theta_{II}}{2}  \ket{01} \, ,
  \nonumber \\
  \ket{d} & = & \sin  \frac{\theta_I}{2}  \ket{00} +
  \cos  \frac{\theta_I}{2}  \ket{11} \, ,
\end{eqnarray}
and correspond respectively to the eigenvalues (ordered by increasing
energies):
\begin{eqnarray}
  \label{eigenvalues}
    E_a  & = & \frac{1}{2}
  (\omega_1 + \omega_2 ) -
  \frac{1}{2} \sqrt{(\omega_2 + \omega_1 )^2 + \lambda^2} \, ,
  \nonumber \\
  E_b  & = &  \frac{1}{2} (\omega_1 + \omega_2 ) -
  \frac{1}{2}\sqrt{(\omega_2 - \omega_1)^2 + \lambda^2} \, ,
  \nonumber \\
    E_c  & = & \frac{1}{2} ( \omega_1 + \omega_2 ) +
  \frac{1}{2}\sqrt{( \omega_2 - \omega_1 )^2 + \lambda^2} \, ,
 \nonumber \\
  E_d  & =  & \frac{1}{2} ( \omega_1 + \omega_2 ) +
  \frac{1}{2}\sqrt{( \omega_2 + \omega_1 )^2 + \lambda^2} \, ,
\end{eqnarray}
where, for $\omega_2 \ge \omega_1$, the parameters $\theta_I$ and
$\theta_{II}$ satisfy the relations
\begin{eqnarray}
  \label{thetaI}
  \displaystyle
  \sin \theta_I & = &  \frac{| \lambda |}
  {\sqrt{( \omega_2 + \omega_1 )^2 + \lambda^2}} \, ,
 \nonumber \\
  \cos \theta_I &  = &  \frac{\omega_1 + \omega_2}
  {\sqrt{( \omega_2  + \omega_1 )^2 + \lambda^2}} \, ,
\nonumber \\
  \sin \theta_{II} &  = & \frac{| \lambda |}
  {\sqrt{(\omega_2 - \omega_1)^2 + \lambda^2}} \, ,
  \nonumber \\
  \cos \theta_{II} & = & \frac{\omega_2-\omega_1}
  {\sqrt{(\omega_2-\omega_1)^2 + \lambda^2}} \, .
\end{eqnarray}
The decay rates $c_i$ and the cross terms $c_{\mathrm{cr},i}$ (with
$i=I,II$) are
\begin{eqnarray}
\label{cI}
  c_I&=&\gamma_{I,11}\left(\cos\frac{\theta_I}{2}\cos\frac{\theta_{II}}{2}
    +\sin\frac{\theta_I}{2}\sin\frac{\theta_{II}}{2}\right)^2\nonumber\\
  \nonumber\\
  & + &\gamma_{I,22}\left(\cos\frac{\theta_I}{2}\sin\frac{\theta_{II}}{2}
    +\sin\frac{\theta_I}{2}\cos\frac{\theta_{II}}{2}\right)^2 \, ,
\nonumber\\
   c_{II}&=&\gamma_{II,11}\left(\cos\frac{\theta_I}{2}\sin\frac{\theta_{II}}{2}
    -\sin\frac{\theta_I}{2}\cos\frac{\theta_{II}}{2}\right)^2
\nonumber\\
 &+&\gamma_{II,22}\left(\cos\frac{\theta_I}{2}\cos\frac{\theta_{II}}{2}
    -\sin\frac{\theta_I}{2}\sin\frac{\theta_{II}}{2}\right)^2 \, ,
\nonumber\\
  c_{cr,I}&=&\gamma_{I,11}\left(\cos\frac{\theta_I}{2}\cos\frac{\theta_{II}}{2}
    +\sin\frac{\theta_I}{2}\sin\frac{\theta_{II}}{2}\right)^2
  \nonumber \\
    &-&\gamma_{I,22}\left(\cos\frac{\theta_I}{2}\sin\frac{\theta_{II}}{2}
    +\sin\frac{\theta_I}{2}\cos\frac{\theta_{II}}{2}\right)^2 \, ,
\nonumber\\
  c_{cr,II}&=&-\gamma_{II,11}\left(\cos\frac{\theta_I}{2}\sin\frac{\theta_{II}}{2}
    -\sin\frac{\theta_I}{2}\cos\frac{\theta_{II}}{2}\right)^2\nonumber\\
  \nonumber\\
  &+&\gamma_{II,22}\left(\cos\frac{\theta_I}{2}\cos\frac{\theta_{II}}{2}
    -\sin\frac{\theta_I}{2}\sin\frac{\theta_{II}}{2}\right)^2,
\end{eqnarray}
where~\cite{petruccionebook}
\begin{eqnarray}
\label{gammasistema}
  \gamma_{i, ll}=J_l(\omega_i) \, \left[1+N_l(\omega_i)\right],
\end{eqnarray}
$J_l(\omega)$ being the zero temperature spectral density of the
l-th reservoir, and $N_l(\omega_i)$  the number of photons in a mode
of frequency $\omega_i$ of the same reservoir. The corresponding
excitation rates $\bar{c}_i$ can be obtained by substituting, to
$\gamma_{i, ll}$ in the corresponding $c_i$, the expression:
\begin{eqnarray}
\label{gammasegnatisistema}
\bar{\gamma}_{i,\,ll} = \gamma_{i,\,ll} \,
\mathrm{e}^{-\omega_i/K_BT_l} =
J_l(\omega_i)\left[N_l(\omega_i)\right] .
\end{eqnarray}
Since $N_l(\omega_i)\rightarrow 0$ for $T_l\rightarrow 0$, the latter
equation clearly shows that the excitation rates vanish at zero
temperature.  We stress that when both reservoir temperatures are
zero, the excitation rates vanish, which physically translates the
{impossibility of creating} excitations in the system due to the
interaction with the reservoirs.

\subsection{Dynamics at zero temperature
and comparison with previous literature}

In reference~\cite{JPHYA} we calculated the two-qubit dynamics at zero
temperatures. {As expected}, all the coherences {oscillate with an
  envelope gradually decreasing to zero}. {As for} the populations,
{the state} $\ket{d}$ decays exponentially towards the states
$\ket{c}$ and $\ket{b}$, which in turn decay towards the ground
state $\ket{a}$.  {From} the time evolution of the populations and
coherences, {the concurrence~\cite{Wootters} can be deduced}. Here,
we focus on two wider classes of initial conditions.

In figure~1.a {the system is} initially prepared in {the}
one-excitation state $\sqrt{p}\ket{01} + \sqrt{1-p}\ket{10}$ and we
plot the concurrence as a function of time and of the weight $p$. In
figure~1.b, {the} system is in a superposition of the states
$\ket{00}$ and $\ket{11}$, namely
$\sqrt{p}\ket{00}+\sqrt{1-p}\ket{11}$.

For the qubit frequencies we take $\omega_1 = \omega_2 =5$GHz, which
are typical values for the current superconducting
technology~\cite{makhlin,noritoday,devoret}. {Since} experiments
indicate that the qubit-qubit coupling constant $\lambda$ can be of
the order of the single qubit frequencies~\cite{7,Nori2009}, we
{take} $\lambda=\omega_1=\omega_2$, in order to maximize the amount
of stationary entanglement~\cite{Ustron}.

\begin{figure}[t]
  \begin{center}
    \includegraphics[width=0.4\textwidth, angle=0]{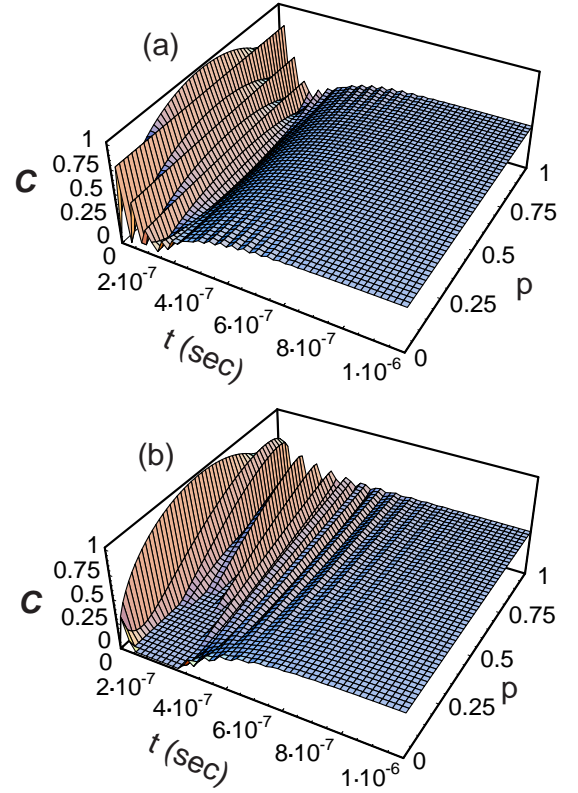}
    \caption{Dynamical evolution of the concurrence, at zero
      temperature, when the system starts from the states (a)
      $\sqrt{p}\ket{01}+\sqrt{1-p}\ket{10}$ and (b) $\sqrt{p}\ket{00}
      + \sqrt{1-p}\ket{11}$ as function of time and of the
      dimensionless weight $p$.}
    \label{figura1}
  \end{center}
\end{figure}

In addition, we consider Ohmic reservoirs with zero-temperature
spectral densities of the form
\begin{eqnarray}
  \label{ohmicJ}
  J_i(\omega)=\alpha_i\,\omega,
\end{eqnarray}
and equal qubit-reservoir coupling strengths $\alpha_1=\alpha_2$.
This implies that, since $\gamma_{I,ii}$ ($\gamma_{{II},ii}$) is
proportional to the spectral density $J_i(\omega_I)$
[$J_i(\omega_{II})$], the ratio between the decay rates is equal to
$c_{II} / c_I = \omega_{II} / \omega_I$. Finally, we assume
$\alpha_1=\alpha_2=10^{-3}\omega_1$ , since it reproduces the decay
times found in recent experiments on superconducting
qubits~\cite{makhlin,noritoday,devoret}.

In figure~1.a Rabi oscillations due to the coherent exchange of
excitations between the two qubits {can be observed at short
  time}. {A different yet oscillatory behavior is seen in
  figure~1.b}, due to the counter-rotating terms in the qubit-qubit
interaction, as well as the phenomena of entanglement sudden
death~\cite{Yu,bellomo,Tanas} and birth~\cite{tanas 2008}, whose
origin can be recognized in the different role of coherences and
populations in appropriately manipulated expressions for the
concurrence.

Concerning the long-time behaviour, both figure~1.a and~1.b show
that {the concurrence reaches a stationary value}. {In
  reference~\cite{tanas 2008}, for example,} the presence of this
entanglement {when there is only a common reservoir} is justified
{in terms of} the very different timescales in the decay of the
symmetric state $(\ket{01}+\ket{10})/\sqrt{2}$ and the antisymmetric
one $(\ket{01}-\ket{10})/\sqrt{2}$. {In fact, this causes the
  strong dependence of the long-time} correlations on the initial
state.

{The situation} in our case is quite different. The origin of the
stationary entanglement can be traced back to the structure of the
eigenstates of the Hamiltonian (\ref{Hamiltonianmodel}). The
zero-temperature dynamics is such that the system eventually goes to
the ground state $\ket{a}$, {which is} a superposition of the
zero-excitation state $\ket{00}$ and the two-excitation state
$\ket{11}$, the latter having a probability amplitude proportional
to the coupling constant $\lambda$, for $\lambda \ll
\omega_1,\omega_2$ [see equation~(\ref{thetaI})]. The net result is
a nonzero stationary value for the concurrence, {irrespective of}
the initial state.

In principle, this independence from the initial state can be
exploited to experimentally discriminate between the entanglement
created by a common reservoir and the one due to the structure of the
ground state, which is a consequence of the presence of
counter-rotating terms in the system Hamiltonian
(\ref{Hamiltonianmodel}). Indeed, if there were no counter-rotating
terms {[that is, a qubit-qubit coupling of the type
  $\lambda(\sigma_-^{(1)}\sigma_+^{(2)}+\sigma_+^{(1)}\sigma_-^{(2)})$]},
the ground state of the system would be {simply} $\ket{00}$, leading
to no stationary entanglement in the absence of a common reservoir
correlating the two qubits.

\section{Dynamics at arbitrary reservoir temperatures}

\subsection{Equal reservoirs}
{Next we} consider the nonzero temperature dynamics.  The solutions
for the coherences given in~\cite{JPHYA} are still valid for
arbitrary reservoir temperatures. On the other hand, from
equation~(\ref{Meqfinale}), it is possible to show that the dynamics
of the populations is governed by the following rate equations:
\begin{eqnarray}
  & &  \dot{\varrho}_{aa} = - (\bar{c}_I + \bar{c}_{II}){\varrho}_{aa} +
  c_I{\varrho}_{bb}+c_{II} {\varrho}_{cc} ,\nonumber\\
  & & \dot{\varrho}_{bb} = \bar{c}_I {\varrho}_{aa} -
  (c_I+\bar{c}_{II}){\varrho}_{bb} + {c}_{II}{\varrho}_{dd}, \nonumber \\
  & & \dot{\varrho}_{cc} = \bar{c}_{II} {\varrho}_{aa} -
  (\bar{c}_I + {c}_{II}) {\varrho}_{cc} + {c}_{I} {\varrho}_{dd} , \nonumber\\
  & & \dot{\varrho}_{dd} = \bar{c}_{II}{\varrho}_{bb} + \bar{c}_I{\varrho}_{cc} -
  ({c}_I +{c}_{II}) {\varrho}_{dd}.
\end{eqnarray}
This can be solved by Laplace transforms, which leads us to the
following algebraic linear equations
\begin{eqnarray}
  & & ( s + \bar{c}_I + \bar{c}_{II} ) \tilde{\varrho}_{aa} -
  c_I \tilde{\varrho}_{bb} - c_{II} \tilde{\varrho}_{cc} = \varrho_{aa}(0) ,
  \nonumber\\
  & & - \bar{c}_I \tilde{\varrho}_{aa} + (s + c_I + \bar{c}_{II} )
  \tilde{\varrho}_{bb} - {c}_{II} \tilde{\varrho}_{dd} = \varrho_{bb}(0),
  \nonumber \\
  & & - \bar{c}_{II} \tilde{\varrho}_{aa} + ( s + \bar{c}_I + {c}_{II} ) \tilde{\varrho}_{cc} - {c}_{I} \tilde{\varrho}_{dd} = \varrho_{cc}(0) ,
  \nonumber \\
  & & - \bar{c}_{II}\tilde{\varrho}_{bb} - \bar{c}_I \tilde{\varrho}_{cc} +
  (s + {c}_I + {c}_{II}) \tilde{\varrho}_{dd} = \varrho_{dd}(0) ,
\end{eqnarray}
where
\begin{eqnarray}
  \tilde{\varrho}_{kk} = \int_{0}^{+\infty} \,  e^{-i s t}
  \varrho_{kk}(t) \,dt.
\end{eqnarray}

After some algebra and transforming back to the time domain, we
finally obtain the  solutions

\newpage

\begin{eqnarray}
  &&\varrho_{aa}(t) =  \frac{c_I c_{II}}{(c_I + \bar{c}_I) (c_{II} +
    \bar{c}_{II})}\\
  &&+ \frac{\bar{c}_I \bar{c}_{II} \varrho_{aa}(0) - {c}_I\bar{c}_{II} \varrho_{bb}(0)
    - \bar{c}_I {c}_{II} \varrho_{cc}(0) + c_Ic_{II}\varrho_{dd}(0)}
  {(c_I+\bar{c}_I)(c_{II}+\bar{c}_{II})}
  \nonumber\\
  &&\times e^{-(c_I+c_{II}+\bar{c}_I+\bar{c}_{II})t} \nonumber\\
  &&+ \frac{{c}_I\bar{c}_{II} [ \varrho_{aa}(0) + \varrho_{bb}(0)]
    - {c}_I {c}_{II} [\varrho_{cc} (0) + \varrho_{dd}(0)]}
  {(c_I + \bar{c}_I) (c_{II} + \bar{c}_{II})} \nonumber\\
  &&\times e^{-(c_{II}+\bar{c}_{II})t} \nonumber\\
  &&+  \frac{\bar{c}_I{c}_{II} [ \varrho_{aa}(0) + \varrho_{cc}(0) ] -
    {c}_I {c}_{II} [ \varrho_{bb} (0) + \varrho_{dd} (0) ]}
  {(c_I + \bar{c}_I) (c_{II} + \bar{c}_{II})}\nonumber\\
  &&\times  e^{-(c_{I}+\bar{c}_{I})t},
  \nonumber \\
  &&\varrho_{bb} (t) =  \frac{\bar{c}_Ic_{II}}{(c_I + \bar{c}_I) (c_{II} +
    \bar{c}_{II})}\nonumber\\
  &&+ \frac{-\bar{c}_I\bar{c}_{II} \varrho_{aa}(0) + {c}_I \bar{c}_{II} \varrho_{bb}(0) +
    \bar{c}_I{c}_{II} \varrho_{cc}(0) - c_Ic_{II} \varrho_{dd}(0)}
  {(c_I + \bar{c}_I) (c_{II} + \bar{c}_{II})}\nonumber\\
  &&\times
  e^{-(c_I+c_{II}+\bar{c}_I+\bar{c}_{II})t} \nonumber \\
   &&+  \frac{\bar{c}_I\bar{c}_{II} [ \varrho_{aa}(0) + \varrho_{bb}(0) ]
    - \bar{c}_I {c}_{II} [ \varrho_{cc} (0) + \varrho_{dd} (0)]}
  {(c_I + \bar{c}_I) (c_{II} + \bar{c}_{II})}
  \nonumber\\
  &&\times  e^{-(c_{II} + \bar{c}_{II}) t} \nonumber\\
  && + \frac{-\bar{c}_I {c}_{II} [ \varrho_{aa}(0) + \varrho_{cc}(0)] +
    {c}_I{c}_{II} [ \varrho_{bb}(0) + \varrho_{dd}(0)]}
  {(c_I + \bar{c}_I) (c_{II} + \bar{c}_{II})}\nonumber\\
  &&\times
  e^{-(c_{I}+\bar{c}_{I})t},
  \nonumber\\
 && \varrho_{cc}(t)  =  \frac{c_I \bar{c}_{II}}{(c_I + \bar{c}_I)
    (c_{II} + \bar{c}_{II})} \nonumber \\
  &&+ \frac{- \bar{c}_I \bar{c}_{II} \varrho_{aa}(0)
    +{c}_I\bar{c}_{II} \varrho_{bb}(0)
    +\bar{c}_I{c}_{II}\varrho_{cc}(0)-c_Ic_{II}\varrho_{dd}(0)}
  {(c_I+\bar{c}_I)(c_{II}+\bar{c}_{II})}
  \nonumber\\
  &&\times e^{-(c_I+c_{II}+\bar{c}_I+\bar{c}_{II})t} \nonumber \\
 && + \frac{-{c}_I \bar{c}_{II} [ \varrho_{aa} (0) + \varrho_{bb}(0)] +
    {c}_I {c}_{II} [ \varrho_{cc}(0) + \varrho_{dd}(0))]}
  {(c_I + \bar{c}_I) (c_{II} + \bar{c}_{II})}
  \nonumber\\
  &&\times e^{-(c_{II}+\bar{c}_{II}) t} \nonumber\\
 && +  \frac{\bar{c}_I \bar{c}_{II} [ \varrho_{aa}(0) + \varrho_{cc}(0)] -
    {c}_I \bar{c}_{II} [ \varrho_{bb}(0) + \varrho_{dd}(0)]}
  {(c_I + \bar{c}_I) (c_{II} + \bar{c}_{II})} \nonumber\\
  &&\times e^{-(c_{I}+\bar{c}_{I})t},
  \nonumber\\
  &&\varrho_{dd}(t)  =  \frac{\bar{c}_I \bar{c}_{II}}
  {(c_I + \bar{c}_I)(c_{II} + \bar{c}_{II})} \nonumber\\
 && +
  \frac{\bar{c}_I \bar{c}_{II} \varrho_{aa}(0) - {c}_I \bar{c}_{II} \varrho_{bb}(0)
    - \bar{c}_I{c}_{II} \varrho_{cc}(0) + c_I c_{II} \varrho_{dd}(0)}
  {(c_I+\bar{c}_I)(c_{II}+\bar{c}_{II})}
  \nonumber\\
  &&\times e^{-(c_I+c_{II}+\bar{c}_I+\bar{c}_{II})t}
  \nonumber\\
  &&+  \frac{- \bar{c}_I \bar{c}_{II} [ \varrho_{aa}(0) + \varrho_{bb}(0)] +
    \bar{c}_I {c}_{II} [ \varrho_{cc}(0) + \varrho_{dd}(0)]}
  {(c_I + \bar{c}_I) (c_{II} + \bar{c}_{II})}
  \nonumber\\
  &&\times e^{-(c_{II}+\bar{c}_{II})t}
  \nonumber\\
  &&+  \frac{- \bar{c}_I \bar{c}_{II} [\varrho_{aa}(0) + \varrho_{cc}(0)] +
    {c}_I \bar{c}_{II} [ \varrho_{bb}(0) + \varrho_{dd}(0)]}
  {(c_I + \bar{c}_I) (c_{II} + \bar{c}_{II})}
  \nonumber\\
  &&\times e^{-(c_{I}+\bar{c}_{I})t},\nonumber
  \label{Popdt}
\end{eqnarray}
which reproduce, at $t=\infty$, the stationary state at generic
temperatures and, for $\bar{c}_I = \bar{c}_{II}= 0$, also the
zero-temperature solutions given in Ref.~\cite{JPHYA}.

\begin{figure}
  \begin{center}
    \includegraphics[width=0.4\textwidth, angle=0]{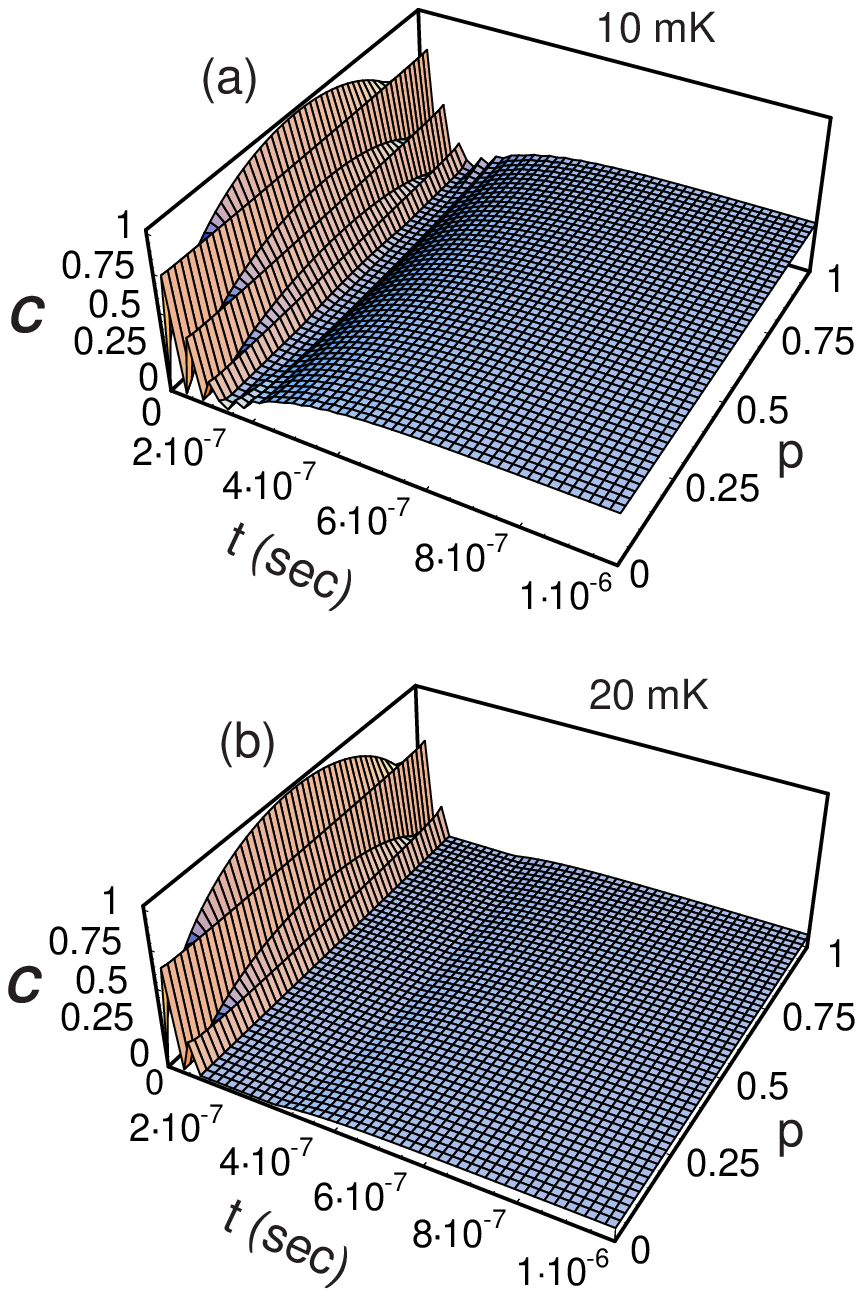}
    \caption{Dynamical evolution of the concurrence, at temperatures
      (a) $T_1=T_2= 10 \,$ mK and (b) $T_1=T_2= 20 \,$ mK, when the
      system starts from the state $
      \sqrt{p}\ket{01}+\sqrt{1-p}\ket{10}$ as function of time and of
      the dimensionless weight $p$.}\label{figura3}
  \end{center}
\end{figure}

\begin{figure}
  \begin{center}
    \includegraphics[width=0.4\textwidth, angle=0]{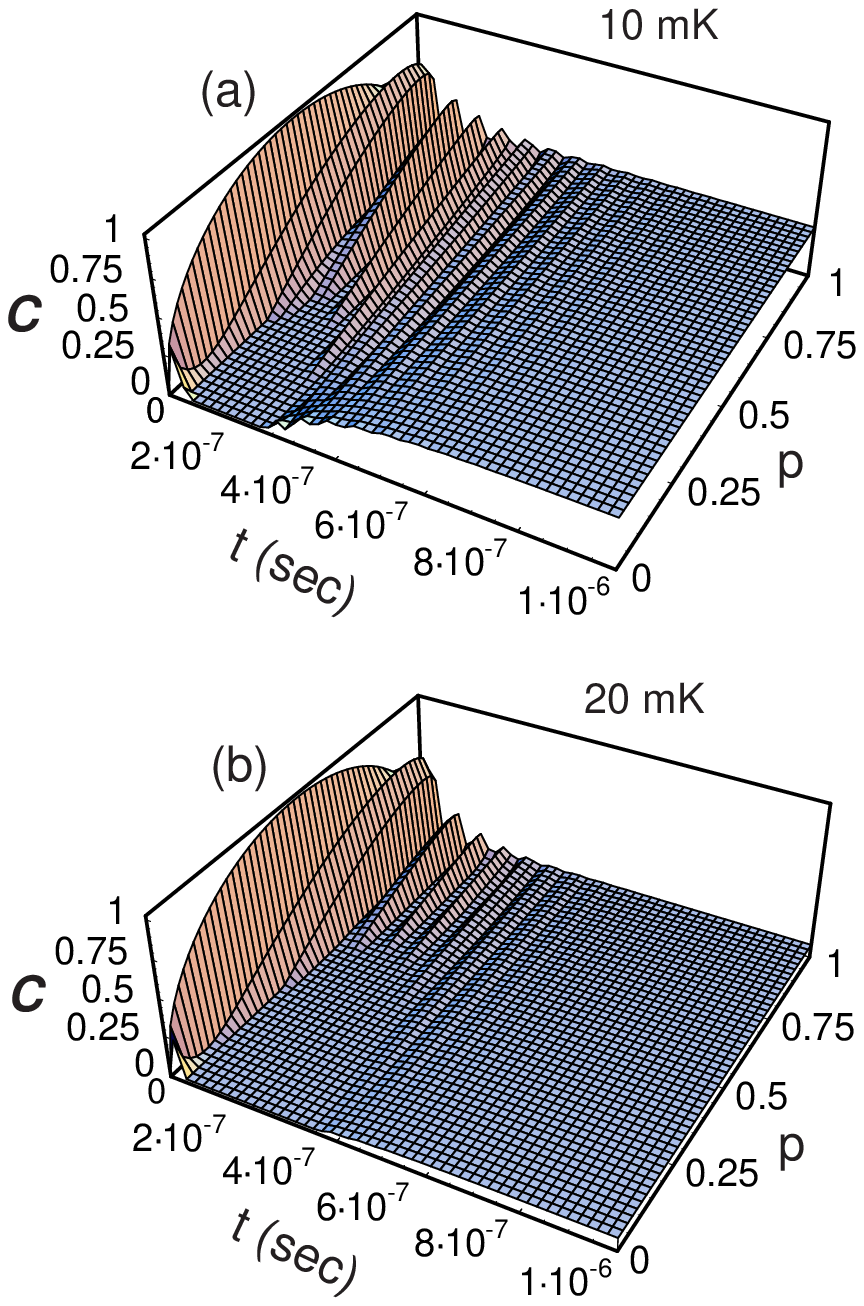}
    \caption{Dynamical evolution of the concurrence, at temperatures
      (a) $T_1=T_2= 10 \,mK$ and (b) $T_1=T_2= 10 \,mK$, when the
      system starts from the state $
      \sqrt{p}\ket{00}+\sqrt{1-p}\ket{11}$ as function of time and of
      the dimensionless weight $p$.}\label{figura4}
  \end{center}
\end{figure}

>From equations~(\ref{Popdt}) it is possible to evaluate the effects of
nonzero reservoir temperatures on the entanglement. In the following
we assume $T_1 = T_2\equiv T$, and again equal coupling strengths
between each qubit and the corresponding reservoir. We postpone to the
next subsection the analysis of the effects of different bath
temperatures and different coupling strengths. Figures~2 and 3 show
the time evolution of the concurrence, as a function of the time $t$
and the weight $p$, when the system starts from the states
$\sqrt{p}\ket{01}+\sqrt{1-p}\ket{10}$ and
$\sqrt{p}\ket{00}+\sqrt{1-p}\ket{11}$, respectively.

{In figure~2 we can see that the oscillations in the
  concurrence}, which mirror the coherent periodic exchange of energy
between the two qubits, are quite robust {for short times} against
thermal noise. Indeed, even at the rather high temperature of 20~mK,
they remain visible for a few Rabi periods.  The same happens to the
short-time qubit-qubit coherent oscillations, caused in this case by
the counter-rotating terms when the system starts from the state
$\sqrt{p}\ket{00}+\sqrt{1-p}\ket{11}$: again the oscillations are
clearly visible at reservoir temperatures of 10 and 20~mK as shown
in figure~3.

{From} both figures~2 and 3 it is apparent that the stationary
entanglement is less robust against thermal noise than the
qubit-qubit coherent dynamics, since at 20 mK it almost disappears.
Anyway its presence can be still detected at temperatures of the
order of 10 mK, which are reasonable values in the case of two
SQUIDs~\cite{makhlin,noritoday,devoret}.

\begin{figure}
  \begin{center}
    \includegraphics[width=0.4\textwidth, angle=0]{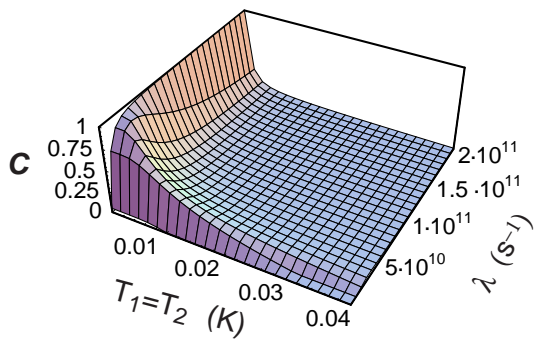}
    \caption{Stationary concurrence $\mathcal{C}(\infty)$ as a
      function of the bath temperatures $T_1=T_2=T$ and the coupling
      constant $\lambda$.  The values of $\lambda$ giving robust
      stationary entanglement are of the order of 10 GHz, i.e. of the
      qubit Bohr frequencies.}
    \label{figura5}
  \end{center}
\end{figure}

{In figure~4 we plot the stationary concurrence}
$\mathcal{C}(\infty)$ as a function of both $T$ and $\lambda$. While
at zero temperature the stationary state is very close to a
maximally entangled state when $\lambda \rightarrow\infty$, for
experimentally meaningful temperatures (i.e., of the order of tens
of mK) the optimal value for stationary concurrence corresponds to
those $\lambda$ up to ten times the qubit Bohr frequencies. Indeed,
for $\lambda \rightarrow\infty$ the value of $\mathcal{C}(\infty)$,
which at zero temperature is almost unity, decays very rapidly with
{temperature}.

\subsection{Different reservoirs}

Let us conclude our analysis by{taking into consideration} different
reservoir temperatures and {qubit-reservoir} coupling strengths.

\begin{figure}[b]
  \begin{center}
    \includegraphics[width=0.4\textwidth, angle=0]{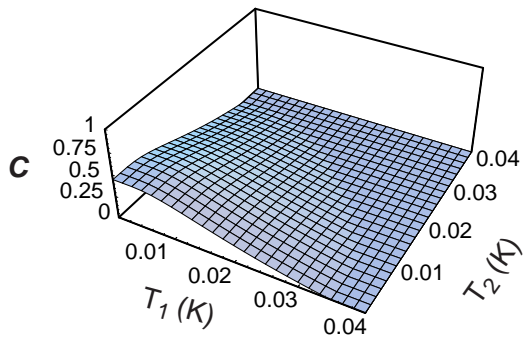}
    \caption{Stationary concurrence $\mathcal{C}(\infty)$ as a
      function of the bath temperatures $T_1$ and $T_2$ for $\alpha_1
      = \alpha_2 = 10^{-3}\omega_1$.}
    \label{figura6}
  \end{center}
\end{figure}

Figure~5 shows the stationary concurrence as a function of the
reservoir temperatures $T_1$ and $T_2$, for
$\lambda=\omega_1=\omega_2$ and equal qubit-reservoir coupling
strengths. Figure~6 compares, in terms of the corresponding contour
plots, this situation with the case in which the coupling strengths
are different.

For equal coupling strengths, we note an almost symmetric role of
the {reservoir temperatures} in the destruction of the stationary
entanglement. From figure~6 we note that the only effect of
different coupling strengths is to erase this symmetry, in the sense
that the larger the coupling strength, the smaller the temperature
necessary to reproduce the same {damping}. This interplay between
temperatures and coupling strengths {can be easily understood} if we
recall that the two reservoirs give independent contributions to the
decay and excitation rates {and the} dynamics is influenced only by
the numerical values of these rates.

\begin{figure}
  \begin{center}
    \includegraphics[width=0.4\textwidth, angle=0]{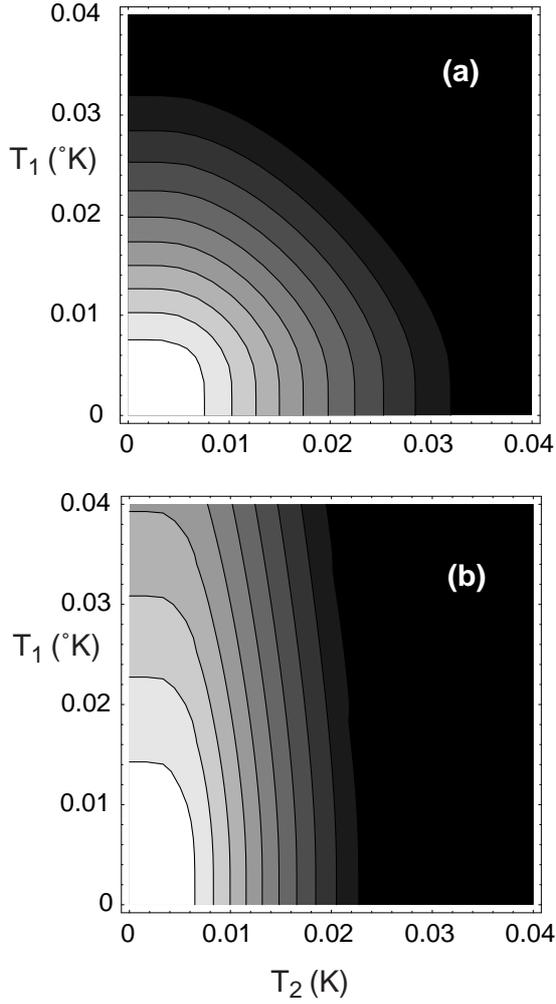}
    \caption{Contour plots of the stationary concurrence
      $\mathcal{C}(\infty)$ as a function of the bath temperatures
      $T_1$ and $T_2$ when (a) $\alpha_1=\alpha_2=10^{-3}\omega_1$ and
      (b) $\alpha_1=10\,\alpha_2=10^{-2}\omega_1$. The white color
      corresponds to $\mathcal{C}(\infty)=0.45$, while the black color
      corresponds to $\mathcal{C}(\infty)=0$.}\label{figura7}
  \end{center}
\end{figure}

\section{Discussion and concluding remarks}

We have presented a general solution of a decay model {for} two
interacting qubits, each one coupled to a {different} bosonic
reservoir, previously solved in the zero-temperature
{only}~\cite{JPHYA}. The analysis reported here clearly shows that
the counter-rotating terms in the qubit-qubit interaction, usually
neglected via a rotating wave approximation, cause two distinct
physical effects, still {observable} at reasonable experimental
temperatures (typically of the order of $10 \div 20$ mK).  First,
{qubit-qubit coherent oscillations can appear even} when the system
starts {in a} superposition of the states $\ket{00}$ and $\ket{11}$,
a case wherein there would be no oscillations in absence of
counter-rotating terms. Second, and more important, the
counter-rotating terms give rise to stationary entanglement, which
is independent from the initial state, {a fact that} can be
exploited to get robust and long-lasting entanglement on demand.

An important point in our analysis is the role of the qubit-qubit
coupling strength $\lambda$ in the creation of the stationary
entanglement. Contrarily to what intuition suggests, it is not
useful to increase $\lambda$ indefinitely: after a certain value,
the stationary entanglement created is very fragile against thermal
noise. Therefore, one can single out a range of values of $\lambda$
within which one can get a reasonable amount of entanglement { even
with } increasing temperatures. The extraction of this robust
entanglement could be useful for {quantum information protocols}.
This point, as well as a non-Markovian extension of the theory
presented here, will be the subject of our future research.

\section*{Acknowledgements}

The authors acknowledge partial support by MIUR Project N.  II04C0E3F3
and DGI Project N. FIS2008-04356. M.S. acknowledges financial support
by the European Commission project EMALI and by the Fondazione Angelo
Della Riccia and wishes to thank Prof. N.  Vitanov from the University
of Sofia, Bulgaria, for useful discussions. The authors also
acknowledge useful discussions with Dr. M.A. Jivulescu.

\end{document}